\documentstyle[prd,aps, 
tighten,amsmath]{revtex}
\begin{document}
\draft
\twocolumn

\title{Decuplet baryon magnetic moments in the chiral quark model}

\author{Johan Linde,\footnote{Email address: jl@theophys.kth.se} Tommy
Ohlsson,\footnote{Email address: tommy@theophys.kth.se} and H{\aa}kan
Snellman\footnote{Email address: snell@theophys.kth.se}} \address{Theoretical
Physics, Department of Physics, Royal Institute of Technology, SE-100
44 Stockholm, Sweden}

\date{Received September 26 1997; revised manuscript received 22
December 1997; published 3 April 1998}

\maketitle

\begin{abstract}
We present calculations of the decuplet baryon magnetic moments in the
chiral quark model.  As input we
use parameters obtained in qualitatively accurate fits to the octet
baryon magnetic moments studied previously. The values found for the
magnetic moments of $\Delta^{++}$ and $\Omega^{-}$ are in good
agreement with experiments.  We finally calculate the total quark spin
polarizations of the decuplet baryons and find that they are considerably
smaller than what is expected from the non-relativistic quark model.
\end{abstract}


\narrowtext

\section{Introduction}	
	
Hadron structure, showing up  in experiments such as deep inelastic
scattering (DIS) on
nucleons, flavor asymmetry measurements in Drell-Yan production, and
measurements of magnetic moments and axial-vector form factors,
does not always fit quantitatively very well with simple
non-relativistic model (NQM) predictions.

It has been difficult, though, to understand these
features from the QCD Lagrangian alone.  Since, at low energies,
phenomena related to chiral symmetry breaking play a major role in
QCD, Manohar and Georgi\cite{mano84} have suggested that these
phenomena can be described by a system of Goldstone bosons (GBs) interacting
with the valence quarks of the NQM. The GBs will give
rise to a polarized quark sea and modify the spin polarizations of the
quarks by the creation of correlated quark-antiquark pairs.  The
relevant scale on which this takes place is $\Lambda_{\chi
\text{SB}}$, which is assumed to be around $1$ GeV. This is much
higher than the confinement scale $\Lambda_{\text{QCD}}$, which is of
the order of 200 MeV. Thus inside hadrons we have a system of quarks,
gluons and GBs interacting with each other.  This
theory, sometimes denoted the chiral quark model ($\chi$QM), has been
used recently to calculate the spin polarization of the quarks in the
proton in DIS \cite{eich92,chen95,lilf95,chen952} and octet baryon magnetic
moments \cite{chen952}. Lately, also SU(3) symmetry breaking in the $\chi$QM
Lagrangian has been included in the calculations
\cite{song97,webe97,chen97,lind97}.

Since the $\chi$QM is quite successful in describing both octet
magnetic moments, the $\bar{u}$-$\bar{d}$ asymmetry, and the spin
polarizations, it is of interest to examine its performance also for
other baryonic systems, such as the spin 3/2 decuplet, using
the same approximation as for the octet baryons.

The decuplet baryon magnetic moments have been calculated in several
models, e.g. in quenched lattice gauge theory \cite{lein92},
quark models \cite{schl93,lind96}, the chiral bag model
\cite{hong94}, chiral perturbation theory \cite{butl94}, chiral
quark-soliton model \cite{kimp97}, and QCD sum rules \cite{lee97,lee972}.

Here we present a calculation of the decuplet baryon
magnetic moments and quark spin polarizations using the $\chi$QM in
the same approximation as for the octet baryons \cite{lind97}.

\section{The Chiral Quark Model}

The GBs of the $\chi$QM are pseudoscalars and
will be denoted by the $0^{-}$ meson names $\pi,K,\eta,\eta'$, as is
usually done.  For convenience we will follow closely the notation of
Refs.  \cite{chen95,lind97}.  The Lagrangian of the interaction is to lowest
order $\tilde{\cal L} = g_8 \bar{\bf q} \tilde{\Phi}\gamma^{5} {\bf q}$ with
\begin{equation}
 	\tilde{\Phi} = \left( \begin{array}{ccc}
 	\frac{\pi^0}{\sqrt{2}}+ \beta \frac{\eta}{\sqrt{6}} & \pi^+ &
 	\alpha K^+ \\ \pi^- & -\frac{\pi^0}{\sqrt{2}}+ \beta
 	\frac{\eta}{\sqrt{6}} & \alpha K^0 \\ \alpha K^- &
 	\alpha\bar{K}^0 & - \beta \frac{2\eta}{\sqrt{6}} \end{array}
 	\right) .  \label{fi}
\end{equation}
We have here introduced two SU(3) symmetry breaking parameters,
$\alpha$ and $\beta$, which allow for different strengths of the
production of GBs containing strange quarks.

In addition to the octet of GBs there is also an SU(3)
singlet of $\eta'$ bosons.  These are coupled to the quarks with
different strength, since the theory would otherwise be U(3) symmetric
(when $\alpha=1$ and $\beta=1$), something that does not agree with
the measurements of the flavor asymmetry by the New Muon
Collaboration (NMC) \cite{amau91,arne94} in DIS and the NA51
Collaboration \cite{bald94} in Drell-Yan production.
The SU(3) scalar interaction has the form
${\cal L}'=g_0 \bar{\bf q} \eta' \gamma^{5}{\bf q} / \sqrt{3}$, so the
total Lagrangian
of interaction is ${\cal L}_I = \tilde{\cal L} + {\cal L}'$.

The effect of this coupling is that the emission of the GBs
will create quark-antiquark pairs from the vacuum with quantum
numbers of the pseudoscalar mesons. Since the GBs are
pseudoscalars, the quark-antiquark fluctuations leave a quark with a spin
opposite to that of the
initial quark, which was absorbed into the GB. This
leads naturally to a spin flip for the quarks. The interaction of the GBs is
weak enough to be treated by perturbation theory. This means that on
long enough time scales for the low energy parameters to develop we
have
\begin{mathletters}\label{1}
\begin{eqnarray}
	u^{\uparrow} & \rightleftharpoons & (d^{\downarrow}+ \pi^+) +
	(s^{\downarrow} +K^+) + (u^{\downarrow} + \pi^0, \eta, \eta')
	, \\ d^{\uparrow}& \rightleftharpoons & (u^{\downarrow}+
	\pi^-) + (s^{\downarrow} +K^0 ) + (d^{\downarrow} + \pi^0
	,\eta ,\eta') , \\ s^{\uparrow} & \rightleftharpoons &
	(u^{\downarrow}+ K^-) + (d^{\downarrow} +\bar{K}^0) +
	(s^{\downarrow} + \eta ,\eta') .
\end{eqnarray}
\end{mathletters}
The probability of transforming a quark with spin up by one
interaction can be expressed by the functions
\begin{mathletters}   \label{2}
\begin{eqnarray}
	\vert\psi(u^\uparrow)\vert^2 &=& \tfrac{1}{6} a (3 + \beta^2 + 2
	\zeta^2) \hat{u}^\downarrow + a \hat{d}^\downarrow + a
	\alpha^{2} \hat{s}^\downarrow, \label{uuppflipp}\\
	\vert\psi(d^\uparrow)\vert^2 &=& a \hat{u}^\downarrow +
	\tfrac{1}{6} a (3 + \beta^2 + 2 \zeta^2) \hat{d}^\downarrow + a
	\alpha^{2} \hat{s}^\downarrow, \\ \vert\psi(s^\uparrow)\vert^2
	&=& a \alpha^{2} \hat{u}^\downarrow + a \alpha^{2}
	\hat{d}^\downarrow + \tfrac{1}{3} a (2 \beta^2 +\zeta^2)
	\hat{s}^\downarrow,
\end{eqnarray}
\end{mathletters}
where the parameter $\zeta \equiv g_0/g_8$.
The coefficient of a quark $\hat{q}^\downarrow$ is the transition
probability to $q^{\downarrow}$.  For example, in Eq.~(\ref{uuppflipp}),
$a(3+\beta^2+2\zeta^{2})/6$ is the probability for
$u^{\uparrow}\to u^{\downarrow}$.  The parameter $a$ is proportional
to $\vert g_8 \vert^2$ and measures the probability of emission of a
GB from a quark.

The total probabilities of no GB emission $P_q$, where $q = u,d,s$, are
\begin{eqnarray}
P_u = P_d &=& 1 - a \left[\left(9 + \beta^2 + 2 \zeta^2\right)/6 + \alpha^2
\right],\\
P_s &=& 1 - a \left[\left(2\beta^2 + \zeta^2\right)/3 + 2\alpha^2\right].
\end{eqnarray}

The spin structure of a baryon $B$ is described by the function
\begin{eqnarray}
{\scriptstyle
	\hat{B}(xyz) = n_{x^{\uparrow}} \hat{x}^{\uparrow} +
	n_{x^{\downarrow}} \hat{x}^{\downarrow} +
	n_{y^{\uparrow}} \hat{y}^{\uparrow} +
	n_{y^{\downarrow}} \hat{y}^{\downarrow} +
	n_{z^{\uparrow}} \hat{z}^{\uparrow} +
	n_{z^{\downarrow}} \hat{z}^{\downarrow}.
}
\end{eqnarray}
The coefficient $n_{q^{\uparrow\downarrow}}$ of each symbol
$\hat{q}^{\uparrow\downarrow}$ is the number of
$q^{\uparrow\downarrow}$ quarks. The spin polarization for a quark $q$
is then defined as
\begin{equation}
	\Delta q^B = n_{q^{\uparrow}}(B)- n_{q^{\downarrow}}(B).
\end{equation}
The magnetic moment of a baryon is now obtained as
\begin{equation}
\mu(B) = \Delta u^B \mu_u + \Delta d^B \mu_d + \Delta s^B \mu_s,
\end{equation}
where $\mu_q$ is the magnetic moment of the quark of flavor
$q$ and $\Delta q^B$ is the corresponding quark spin polarization for
the baryon $B$.
The total quark spin polarization of a baryon B is given by the expression
\begin{equation}
\Delta \Sigma^B = \Delta u^B + \Delta d^B + \Delta s^B.
\label{eq:totspinpol}
\end{equation}

The expressions for the quark spin polarizations for the octet
baryons can be found in Ref. \cite{lind97}.

\section{Octet Baryons}

First we have to specify the values of the parameters used in the
model.

The constants $a$ and $\zeta$ are estimated from the $\bar
u$-$\bar d$ asymmetry ($\bar{u} - \bar{d} = - 0.15 \pm 0.04$ and
$\bar{u}/\bar{d} = 0.51 \pm 0.09$). When $\beta=1$, the value of
$\zeta$ is found to be in the interval $-4.3 < \zeta < -0.7$.
Following Cheng and Li\cite{chen95}, we use the value $\zeta = -1.2$,
which gives $a \approx 0.10$. ($a \approx 0.10$ is in good agreement
with Ref. \cite{eich92}.) However, when $\beta$ is a free
parameter in the calculations, the relation between $a$, $\zeta$, and
$\beta$ is
\begin{equation}
a \simeq 0.45 / (3-2 \zeta-\beta).
\label{zeta}
\end{equation}
This means that we have to use the relation $2 \zeta + \beta \simeq
-1.4$, in order to keep $a \approx 0.10$, and therefore $\zeta = - 0.7
- \beta/2$.

We also use the relations $\mu_u = - 2 \mu_d$ and $\mu_s = 2 \mu_d
/3$, typically used in the NQM.

The remaining three parameters $\alpha$, $\beta$, and $\mu_d$
are determined by a fit to the
magnetic moments of the octet baryons \cite{lind97}.
As a result from this fit, we obtained $\alpha \approx 0.52$,
$\beta \approx 0.99$, and $\mu_{d} \approx -1.23 \, \mu_N$.

Recently, other choices of parameters have also been attempted
\cite{webe97,chen97}. One such choice is $\zeta=-0.3$.
For $\alpha=\beta=0.6$ this simplifies
Eq.~(\ref{zeta}), so that $a=0.15\pm 0.04$. However, the
value of $\bar u/\bar d$ becomes quite high, $0.68$.

Since our fits affirm that $\beta \approx 1$, we have fixed the value of this
parameter to $\beta=1$ in a fit with $\zeta=-0.5$, which is presented for
comparison. This value corresponds to no suppression of the $\eta$ contribution
$(\beta=1)$ and large $(\zeta =-0.5)$ suppression of the $\eta'$ contribution.
The price to be paid is that the value for  $\bar u/\bar d$ goes out of
the experimental range and becomes $0.64$. Since this measurement is made
only for $x=0.18$, this might nevertheless be a possibility to keep in mind.

Our parameters $\alpha$ and $\beta$ correspond to the
parameters $\epsilon$ and $\delta$, respectively, in Ref. \cite{chen97},
whereas the parameter $\epsilon$ used in Ref. \cite{webe97} is related to both
our parameters $\alpha$ and $\beta$. The value $\epsilon=0.2$ used
by \cite{webe97} corresponds to our parameter values $\alpha \approx 0.77$ and
$\beta\approx 1.86$. The high value for $\beta$, when $a=0.12$, which keeps the
proton quark spin polarization down, destroys the agreement with the magnetic
moments (which prefer $\beta \approx 1$). We have therefore kept the
value $\beta=1$, and fixed $a$ to $0.15$.

One could try to fix $\zeta=0$, corresponding to complete suppression of
the $\eta'$ contribution, and let $\beta=1$, corresponding to no $\eta$
suppression. However, the value of $\bar u/\bar d$ then becomes $0.75$
with $\bar u -\bar d \approx
-0.08$. We have therefore not included this case in the tables.

Finally, we have also made a fit with only $\mu_{d}$ as a
free parameter, i.e. when we have put $\alpha = \beta = 1$ and
$\zeta=-1.2$. See Table~\ref{tab:fit_data} for the result.

\section{Decuplet Baryons}

The $\chi$QM can easily be applied to the decuplet. This will give us
a further test of the predictability of the $\chi$QM and also a
possibility to estimate the fraction of the spin carried by the quarks
in the decuplet baryons. We will use the $\chi$QM in the same
approximation for the decuplet as for the octet. The quark masses used in
these fits should be understood as effective quark masses. They
incorporate, together with the parameter $a$, describing the emission
probability of the GBs, effects of relativistic corrections and
possible exchange currents \cite{dann97}. When relativistic corrections are
included explicitly in the model, the fits become worse \cite{webe972}.

In order to calculate the spin polarization for the quarks
in $\Delta^{++}$, we must
count the number of quarks of each flavor in the two polarizations states.

For example, the probability of transforming a $u^\uparrow$ quark with
one interaction is given by Eq.~(\ref{uuppflipp}).  After one
interaction, all quarks that have participated in this interaction have
their spins down.  For a particle like $\Delta^{++}(J_z=+3/2)$, which
has all valence $u$ quarks with spin up, we can in this way understand
that it will create a quite negatively polarized sea around it in the
$\chi$QM.

The general expression for  the spin structure of the decuplet
baryons
$B(xxy)$ is after one interaction  given by
\begin{equation}
\hat{B}(xxy) = 2 P_x \hat{x}^\uparrow + P_y \hat{y}^\uparrow
+ 2 \vert \psi(x^\uparrow) \vert^2 + \vert \psi(y^\uparrow) \vert^2.
\end{equation}
Similarly, the spin structure for the $\Sigma^{\ast 0}$ is given by
\begin{eqnarray}
\hat{\Sigma}^{\ast 0}(uds) &=& P_u \hat{u}^\uparrow
+ P_d \hat{d}^\uparrow + P_s \hat{s}^\uparrow \nonumber \\
&+&
\vert \psi(u^\uparrow) \vert^2 + \vert \psi(d^\uparrow) \vert^2
+ \vert \psi(s^\uparrow) \vert^2.
\end{eqnarray}

The spin structures now lead to the following quark spin polarizations.

For the $\Delta^{++}$ the spin polarizations are given by
\begin{eqnarray}
\Delta u^{\Delta^{++}} &=& 3 - a \left( 6 + 3 \alpha^2 + \beta^2 + 2 \zeta^2
\right), \label{eq:d_begin}\\
\Delta d^{\Delta^{++}} &=& - 3 a, \\
\Delta s^{\Delta^{++}} &=& - 3 a \alpha^2,
\end{eqnarray}
and for the $\Delta^+$ by
\begin{eqnarray}
\Delta u^{\Delta^+} &=& 2 - a \left( 5 + 2 \alpha^2 + \tfrac{2}{3} \beta^2 +
\tfrac{4}{3} \zeta^2 \right), \\
\Delta d^{\Delta^+} &=& 1 - a \left( 4 + \alpha^2 + \tfrac{1}{3} \beta^2 +
\tfrac{2}{3} \zeta^2 \right), \\
\Delta s^{\Delta^+} &=& - 3 a \alpha^2,
\end{eqnarray}
and for the $\Sigma^{\ast +}$ by
\begin{eqnarray}
\Delta u^{\Sigma^{\ast +}} &=& 2 - a \left( 4 + 3 \alpha^2 + \tfrac{2}{3}
\beta^2 + \tfrac{4}{3} \zeta^2 \right), \\
\Delta d^{\Sigma^{\ast +}} &=& - a \left( 2 + \alpha^2 \right), \\
\Delta s^{\Sigma^{\ast +}} &=& 1 - a \left( 4 \alpha^2 + \tfrac{4}{3} \beta^2 +
\tfrac{2}{3} \zeta^2 \right),
\end{eqnarray}
and for the $\Sigma^{\ast 0}$ by
\begin{eqnarray}
\Delta u^{\Sigma^{\ast 0}} &=& 1 - a \left( 3 + 2 \alpha^2 + \tfrac{1}{3}
\beta^2 + \tfrac{2}{3} \zeta^2 \right), \\
\Delta d^{\Sigma^{\ast 0}} &=& 1 - a \left( 3 + 2 \alpha^2 +
\tfrac{1}{3} \beta^2 + \tfrac{2}{3} \zeta^2 \right), \\
\Delta s^{\Sigma^{\ast 0}} &=& 1 - a \left( 4 \alpha^2 + \tfrac{4}{3} \beta^2 +
\tfrac{2}{3} \zeta^2 \right),
\end{eqnarray}
and for the $\Xi^{\ast 0}$ by
\begin{eqnarray}
\Delta u^{\Xi^{\ast 0}} &=& 1 - a \left( 2 + 3 \alpha^2 + \tfrac{1}{3} \beta^2
+ \tfrac{2}{3} \zeta^2 \right), \\
\Delta d^{\Xi^{\ast 0}} &=& - a \left( 1 + 2 \alpha^2 \right), \\
\Delta s^{\Xi^{\ast 0}} &=& 2 - a \left( 5 \alpha^2 + \tfrac{8}{3} \beta^2 +
\tfrac{4}{3} \zeta^2 \right),
\end{eqnarray}
and for the $\Omega^-$ by
\begin{eqnarray}
\Delta u^{\Omega^-} &=& - 3 a \alpha^2, \\
\Delta d^{\Omega^-} &=& - 3 a \alpha^2, \\
\Delta s^{\Omega^-} &=& 3 - a \left( 6 \alpha^2 + 4 \beta^2 + 2
\zeta^2 \right).
\label{eq:d_end}
\end{eqnarray}

The spin polarizations for the other decuplet baryons are found from
isospin symmetry.

We now use the parameters found for the octet baryons to calculate
the magnetic moments and the total quark spin polarizations of the decuplet
baryons.  The results are presented in Tables~\ref{tab:decuplet} and
\ref{tab:spinpol}.

In Table~\ref{tab:decuplet} we present the results obtained for the
magnetic moments of the decuplet baryons. We observe that the value
of $\mu(\Omega^-)$ in the $\chi$QM and the $\chi$QM with SU(3)
symmetry breaking are closer
to experiments than the NQM value, while the values for $\mu(\Delta^{++})$ are
almost equivalent. This is quite non-trivial, since the quark spin
polarizations in the $\chi$QM deviate substantially from the NQM values.

In Table~\ref{tab:spinpol} we list the total quark spin polarizations for
the decuplet baryons. Using Eq.~(\ref{eq:totspinpol}) together with
Eqs.~(\ref{eq:d_begin}) -- (\ref{eq:d_end}), we obtain the total quark spin
polarization for a baryon $B$ as
\begin{equation}
{\scriptstyle \Delta \Sigma^B = 3 - a \left( 9 + 6 \alpha^2 + \beta^2
+ 2 \zeta^2 \right) + a \left( 3 - 2 \alpha^2 - \beta^2
\right) x,}
\label{deltasigma}
\end{equation}
where $x$ is the number of $s$ quarks in the baryon $B$.
Note that in both the NQM and the $\chi$QM with
no SU(3) symmetry breaking, the $\Delta \Sigma$'s have the same value
for all decuplet baryons, but when introducing SU(3) symmetry
breaking, they vary linearly with the number of strange quarks.

For the octet baryons there is no formula similar to Eq.~(\ref{deltasigma}).

\section{Summary and Conclusions}

Applying the results from calculations of the octet baryons
to the decuplet baryons, we have found values for
the magnetic moments of $\Delta^{++}$ and $\Omega^{-}$ in good
agreement with experiments.  This holds not only for the parameters
used in the previous fits, but also for some other choices of the SU(3)
breaking parameters.

The case $\zeta=-0.5$ represents a large $\eta'$ suppression. However, one has
to pay the price that the $\bar u/\bar d$ asymmetry becomes $0.64$.
Fixing the value of $a$ to $0.15$ brings the quark spin polarization
of the proton down to $0.34$. On the other hand, the octet magnetic moments
are then not quite as good.

In our opinion, the quantities $\Delta \Sigma^{p}$,
$\bar u/\bar d$, etc., often used to determine
the parameters of the $\chi$QM are not better understood, and
certainly less well measured, than the magnetic moments chosen here to fix
the parameters, and therefore subject to at least as much uncertainty.
To ask for complete agreement already at this
level for each item might be to ask for too much.

The fraction of the baryon spin carried by the quarks in the decuplet
baryons is in our analysis predicted to be
about one third of the value from the NQM when there is no SU(3) symmetry
breaking and to be slightly larger and increase with the number of strange
quarks
when SU(3) is broken.

\bigbreak

This work was supported by the Swedish Natural Science Research
Council (NFR), Contract No.  F-AA/FU03281-310.
Support for this work was also provided by the Ernst Johnson
Foundation (T.O.).

\begin{table}
\caption{Parameter values obtained in the different fits.  The values
in columns 3, 4, and 6 can be found in Ref. \protect\cite{lind97}.
Hyphen (-)
indicates that the parameter was not defined in the fit. The
magnetic moment of the $d$ quark, $\mu_d$, is given in units of the
nuclear magneton, $\mu_N$.}
\begin{tabular}{ccrrrr}
Parameter/ & Experimental & NQM & $\chi$QM & \multicolumn{2}{c}{$\chi$QM
with SU(3)} \\
Quantity & value & &  &\multicolumn{2}{c}{{\footnotesize symmetry breaking}}\\
$\zeta$ & & & $ -1.2$ & $ -0.5$ & $ -1.2$\\
\hline
$\mu_d$ & & $-0.91$ & $-1.35$ & $-1.30$ & $-1.23$\\
$a$ & & - & $0.10$\tablenotemark[1] & 0.15\tablenotemark[1] &
$0.10$\tablenotemark[1]\\
$\alpha$ & & - & $1$\tablenotemark[1] & $0.67$ & $0.52$\\
$\beta$ & & - & $1$\tablenotemark[1] & $1$\tablenotemark[1] & $0.99$\\
\hline
$\bar{u} - \bar{d}$ & $-0.15 \pm 0.04$ & - & $-0.15$ &  $-0.15$ & $-0.15$\\
$\bar{u}/\bar{d}$ & $0.51 \pm 0.09$ & - & $0.53$ &  $0.64$ &
$0.53$\\
$g_A$ & $1.26 \pm 0.01$ & $\frac{5}{3}$ & $1.12$ &   $1.18$ & $1.24$\\
$\Delta \Sigma^p$ & $0.30 \pm 0.06$ & $1$ & $0.37$ &  $0.34$ & $0.52$\\
\end{tabular}
\tablenotetext[1]{Fixed in fit.}
\label{tab:fit_data}
\end{table}

\begin{table}
\caption{Decuplet magnetic moments. The
decuplet magnetic moments are given in units of the nuclear magneton,
$\mu_N$. In the Review of Particle Physics\protect\cite{barn96} one
can find the averages $\mu(\Delta^{++})= (3.7 \to 7.5)\mu_N$ and
$\mu(\Omega^-)=(-2.02 \pm 0.05)\mu_N$ obtained from all existing
experiments.}
\begin{tabular}{lcddrr}
{\footnotesize Quantity} & {\footnotesize Experimental} & NQM & $\chi$QM &
\multicolumn{2}{c}{$\chi$QM {\footnotesize with} SU(3) }\\
& {\footnotesize value}& & &\multicolumn{2}{c}{{\footnotesize symmetry
breaking}}\\
$\zeta$ & & & $-$1.2 & $-$0.5 & $-$1.2\\
\hline
$\mu(\Delta^{++})$ & $4.52 \pm 0.95${\scriptsize  \protect\cite{boss91}}
& 5.43 & 5.30 &  5.11 & 5.21\\
$\mu(\Delta^+)$ & - & 2.72 & 2.58 &  2.35 & 2.45\\
$\mu(\Delta^0)$ & - & 0 & $-$0.13 &  $-$0.41 & $-$0.30\\
$\mu(\Delta^-)$ & - & $-$2.72 & $-$2.85 &  $-$3.17 & $-$3.06\\
$\mu(\Sigma^{\ast +})$ & - & 3.02 & 2.88 &  2.77 & 2.85\\
$\mu(\Sigma^{\ast 0})$ & - & 0.30 & 0.17 &  0.00 & 0.09\\
$\mu(\Sigma^{\ast -})$ & - & $-$2.41 & $-$2.55 &  $-$2.76 &
$-$2.66\\
$\mu(\Xi^{\ast 0})$ & - & 0.60 & 0.47 &  0.42 & 0.49\\
$\mu(\Xi^{\ast -})$ & - & $-$2.11 & $-$2.25 &  $-$2.34 &
$-$2.27\\
$\mu(\Omega^-)$ & $-1.94 \pm 0.31$ {\scriptsize \protect\cite{dieh91} }
&
$-$1.81
& $-$1.95 & $-$1.93 & $-$1.87\\
& $-2.02 \pm 0.06$ {\scriptsize \protect\cite{wall95}}
\\
\end{tabular}
\label{tab:decuplet}
\end{table}

\begin{table}
\caption{Total quark spin polarizations of the decuplet baryons.}
\begin{tabular}{lddrr}
{\footnotesize Quantity} & NQM & $\chi$QM & \multicolumn{2}{c}{$\chi$QM
{\footnotesize with} SU(3)}\\
 & & & \multicolumn{2}{c}{{\footnotesize symmetry breaking}}\\
 $\zeta$ & &   $-$1.2  &   $-$0.5 & $-$1.2\\
\hline
$\Delta \Sigma^\Delta$ & 3 & 1.11 &  1.03 & 1.55\\
$\Delta \Sigma^{\Sigma^\ast}$ & 3 & 1.11 & 1.19 & 1.70\\
$\Delta \Sigma^{\Xi^\ast}$ & 3 & 1.11 & 1.36 & 1.85\\
$\Delta \Sigma^{\Omega}$ & 3 & 1.11 &  1.53 & 2.00\\
\end{tabular}
\label{tab:spinpol}
\end{table}

\end{document}